\documentclass[aps,prb,twocolumn,superscriptaddress,showpacs,amsmath,amssymb,footinbib,longbibliography]{revtex4-1}
\usepackage[english]{babel}
\usepackage{amssymb,amsbsy,amsmath}
\usepackage{graphicx}
\usepackage{dcolumn}
\usepackage{bm}
\usepackage{subfigure}
\usepackage{color}
\usepackage{textcomp}
\usepackage[colorlinks,bookmarks=false,citecolor=blue,linkcolor=red,urlcolor=blue]{hyperref}

\newcommand{\op}[1]{%
    \fontdimen12\textfont3=2pt\fontdimen12\scriptfont3=1.4pt%
    \!\null\mathop{\vphantom{#1}\smash{#1}}\limits_{\sim}\null\!}

\newcommand{\figref}[1]{Fig.~\protect\ref{#1}}

\def\ket#1{\, | \, {#1} \, \rangle}
\newcommand{\braket}[2]{\langle \, {#1} \, | \, {#2} \, \rangle}

\newcommand{\kagome}{kagom\'e}
\begin{document}
\title{Magnetism of the $N=42$ \kagome\ lattice antiferromagnet}

\author{J\"urgen Schnack}
\email{jschnack@uni-bielefeld.de}
\affiliation{Fakult\"at f\"ur Physik, Universit\"at Bielefeld, Postfach 100131, D-33501 Bielefeld, Germany}
\author{J\"org Schulenburg}
\affiliation{Universit\"atsrechenzentrum, Universit\"at Magdeburg, D-39016 Magdeburg, Germany}
\author{Johannes Richter}
\email{Johannes.Richter@physik.uni-magdeburg.de}
\affiliation{Institut f\"ur Theoretische Physik, Universit\"at Magdeburg, P.O. Box 4120, D-39016 Magdeburg, Germany}
\affiliation{Max-Planck-Institut f\"{u}r Physik Komplexer Systeme,
        N\"{o}thnitzer Stra{\ss}e 38, 01187 Dresden, Germany}

\date{\today}

\begin{abstract}
For the paradigmatic frustrated spin-half Heisenberg antiferromagnet on
the \kagome\ lattice we performed large-scale numerical
investigation of thermodynamic functions by means of the
finite-temperature Lanczos method for system sizes of up to
$N=42$. We present the dependence of magnetization as well as
specific heat on temperature and external field and show
in particular that a finite-size scaling of specific heat
supports the appearance of a low-temperature shoulder below
the major maximum. This seems to be the result of a
counterintuitive motion of the density of singlet states towards
higher energies. Other interesting features that we discuss are
the asymmetric melting of the 1/3 magnetization plateau as well the
field dependence of the specific heat that exhibits characteristic features
caused by the existence of a flat one-magnon band.  
By comparison with the unfrustrated square-lattice antiferromagnet
the tremendous role of
frustration in a wide temperature range is illustrated.
\end{abstract}


\maketitle

\section{Introduction}

The spin-$1/2$ \kagome\ Heisenberg antiferromagnet (KHAF) 
is one of the most prominent and at the same time challenging spin
models in the  field of frustrated quantum magnetism.
The first challenge concerns the nature of the ground state
(GS), on which a plethora of studies
exist, see, e.g., Refs.~\onlinecite{Elser1990,Waldtmann1998,AHL:PRL98,Yu2000,Bernhard2002,Singh2007, 
DMRG_PRL08,Laeuchli2009,Evenbly2010,Goetze2011,Nakano2011,Iqbal2011,
Yan2011,Laeuchli2011,DMS:PRL12,Rousochatzakis2013,
Iqbal2013,Rousochatzakis2014,Xie2014,Kolley2015,
Goetze2015,Laeuchli2016,Pollmann2017,Xie2017,Mei2017,Xi-Chen2017,NaS:JPSJ18}. 
Although consensus on the absence of magnetic long-range order 
(LRO) is achieved, the precise nature of the spin-liquid GS, with quantum spin liquids and Dirac spin liquids as
possible candidates,\cite{Nor:RMP16,Pollmann2017} is
not yet understood. Large-scale density matrix renormalization
group (DMRG) and exact diagonalization (ED) studies
\cite{Yan2011,DMS:PRL12,Laeuchli2011,Laeuchli2016,Pollmann2017} suggest
a  tiny singlet-singlet gap $\Delta_s \sim (0.01 \ldots  0.05)
J$, where $J$ denotes the exchange coupling in the Heisenberg model,
and a sizeable singlet-triplet gap  $\Delta_t \sim (0.13 \ldots
0.17) J$.
However, a very recent DMRG study using adiabatic flux insertion provides
indications for a much smaller spin-gap in agreement with variational and
other numerical techniques.\cite{Iqbal2011,Iqbal2013,Pollmann2017,Xie2017,Xi-Chen2017}
The very existence of a gap is determinative for thermodynamics
at low temperatures $T$. In addition, a triplet gap leads to an
exponentionally activated low-temperature behavior of the 
susceptibility. On the other hand, indications were found that   
a huge number of singlet states below the first triplet state
may exist,\cite{AHL:PRL98,Waldtmann1998,Lhuillier_thermo_PRL2000,DMRG_PRL08,Laeuchli2011,Yan2011,Rousochatzakis2013,Laeuchli2016} 
being relevant for the specific heat $C$ at low temperatures.

Besides the theoretical work there is also large activity on
the experimental side, see, e.g., Refs.
\onlinecite{Atw:NM02,KNO:JPSJ04,HMS:PRL10,RTT:CM11,YMT:JMC12,
herbertsmithite2007,herbertsmithite2007a,Hiroi2009,
herbertsmithite2009,herbertsmithite2010,AAG:NC11,herbertsmithite2012,FIH:S15,FLM:CPL17,GKZ:PRL17,OBM:PRL17,ZHG:PRL17}
and in particular the review~\onlinecite{Nor:RMP16}.
Among the spin-$1/2$ \kagome\ compounds, Herbertsmithite
ZnCu$_3$(OH)$_6$Cl$_2$ seems to be a promising candidate for a
spin liquid.\cite{herbertsmithite2007,herbertsmithite2007a,
Hiroi2009,herbertsmithite2009,herbertsmithite2010,herbertsmithite2012,CHF:PRL15,Nor:RMP16,ZHG:PRL17}

The second challenge concerns the thermodynamic properties of
the quantum KHAF on which far less studies
exist.\cite{elstner1994,NaM:PRB95,ToR:PRB96,Lhuillier_thermo_PRL2000,Yu2000, 
Bernhard2002,Bernu2005,MiS:EPJB07,RBS:PRE07,Lohmann2014,
Mun:WJCMP14,Singh2017,Shimokawa2016,Xi-Chen2017,Misawa2018,kagome-RGM2018}
While systematic high-temperature approaches
\cite{elstner1994,RBS:PRE07,Lohmann2014,Singh2017} 
provide reliable insight into the temperature dependence of physical
quantities down to temperatures $T$ of about 40\% of the
exchange coupling $J$, 
a reliable picture of the temperature dependences at $0 \le T
\lesssim 0.4J$ is still missing. Various methods 
\cite{elstner1994,NaM:PRB95,ToR:PRB96,Bernu2005,Mun:WJCMP14,Shimokawa2016,Misawa2018}
provide indications for an additional low-temperature peak 
of the specific heat signaling 
an extra low-energy scale set by low-lying singlets.
However, instead of a true maximum a shoulder-like hump may characterize the
low-$T$ profile of $C(T)$.\cite{Bernu2005,Xi-Chen2017}
Thus, the low-$T$ behavior of the specific heat is another issue (in some
relation to the gaps) that is not yet settled.

The third challenge is given by the magnetization process of the
spin-1/2
KHAF.\cite{Hida2001,SHS:PRL02,HSR:JP04,HCG:PB05,SaN:PRB11,
Nishimoto2013,CDH:PRB13,NaS:JPSJ14,Xi-Chen2017,PMH:X18,NaS:JPSJ18} 
A series of magnetization plateaus at $3/9(=1/3)$, $5/9$ and $7/9$ of the
saturation is found,\cite{Nishimoto2013,CDH:PRB13,Xi-Chen2017} among which
the $1/3$--plateau, already found by Hida,\cite{Hida2001} is the widest.
In addition to the above mentioned plateaus, there might be a tiny plateau at
$1/9$.\cite{Nishimoto2013,Xi-Chen2017}
The magnetic ordering within the plateau is well-described by
valence-bond states, i.e., the plateau states are of quantum
nature.\cite{HCG:PB05,Nishimoto2013,CDH:PRB13} Moreover, there
is a macroscopic jump to saturation related to the existence of
a huge manifold of localized multi-magnon
states.\cite{SHS:PRL02,derzhko2004finite,zhitomirsky2004exact,DRH:LTP07} 
At low enough temperatures and for specific values of the
magnetic field the plateaus as well as the magnetization jump
are well expressed features of the magnetization curve.
From the experimental point of view the detection of these
features at low temperatures provides \emph{smoking gun}
evidence of the proximity of the investigated  magnetic \kagome\
compound to the ideal KHAF model.      

In the present paper we discuss the thermodynamic properties of the spin-1/2
KHAF on a finite lattice of $N=42$ sites. These results were
obtained by large-scale numerical calculations 
(5 Mio. core hours) using the
finite-temperature Lanczos method
(FTLM).\cite{PhysRevB.49.5065,JaP:AP00,ScW:EPJB10,HaS:EPJB14,ScT:PR17}  
The extension to a lattice of this size yields an improved
insight into the low-temperature physics 
of the model compared to previous ED and FTLM studies restricted
to significantly smaller lattices.

The paper is organized as follow. In Section \ref{sec-2} we
introduce the model and our numerical scheme. Thereafter in
Section~\ref{sec-3} we present our results for the KHAF followed
by a discussion in Section~\ref{sec-4}.

\section{Hamiltonian and calculational scheme}
\label{sec-2}

The investigated spin systems are modeled by a spin-$1/2$
Heisenberg Hamiltonian augmented with a Zeeman term, i. e.
\begin{eqnarray}
\label{E-2-1}
\op{H}
&=&
J\;
\sum_{<i,j>}\;
\op{\vec{s}}_i \cdot \op{\vec{s}}_j
+
g \mu_B\, B\,
\sum_{i}\;
\op{s}^z_i
\ .
\end{eqnarray}
Quantum mechanical operators are marked by a tilde. 
In what follows we set the antiferromagnetic nearest-neighbor
exchange coupling to $J=1$. The complete
eigenvalue spectrum of a spin system composed of spins $s=1/2$
can be evaluated for sizes of up to about $N=24$ depending on
the available symmetries.\cite{ScS:IRPC10} The resulting
thermodynamic quantities are then numerically exact. 

For larger systems with Hilbert
space dimensions of up to $10^{10}$ FTLM provides approximations
of thermodynamic functions with astonishing
accuracy.\cite{ScW:EPJB10,HaS:EPJB14,ScT:PR17} FTLM approximates
the  partition function in two
ways:\cite{PhysRevB.49.5065,JaP:AP00} 
\begin{eqnarray}
\label{E-2-3}
Z(T,B)
&\approx&
\sum_{\gamma=1}^\Gamma\;
\frac{\text{dim}({\mathcal H}(\gamma))}{R}
\sum_{\nu=1}^R\;
\sum_{n=1}^{N_L}\;
e^{-\beta \epsilon_n^{(\nu)}} |\braket{n(\nu)}{\nu}|^2
\ .
\nonumber \\[-3mm]
\end{eqnarray}
The trace, i.e., the sum over an orthonormal basis,
is in a Monte-Carlo fashion replaced by a much smaller sum over
$R$ random vectors 
$\ket{\nu}$ for each symmetry-related orthogonal subspace
${\mathcal H}(\gamma)$ of the Hilbert space.
The exponential of the Hamiltonian is then approximated by
its spectral representation in a Krylov space spanned by the
$N_L$ Lanczos vectors starting from the respective random vector
$\ket{\nu}$. $\ket{n(\nu)}$ is the n-th eigenvector of $\op{H}$ in
this Krylov space. This allows to evaluate typical observables
such as magnetization and specific heat.\cite{PRE:COR17}

The method was implemented in two independently self-written
programs, one of which -- \verb§spinpack§ -- is publicly
available.\cite{spin:256} The latter employs several symmetries
in order to decompose the full Hilbert space into much smaller
orthogonal subspaces according to the irreducible
representations of the used symmetries. In our case $\op{S}^z$
symmetry was used together with the longest cyclic point group
(length 14 for $N=42$) as well as with spin-flip-symmetry and a
second commuting 
point group where applicable. 
The largest Hilbert sub-spaces in the sector with
magnetic quantum number $M=1$ assumed a dimension of 
$3.67\cdot 10^{10}$. We used $R=10$ in all subspaces of $M=0$,
i.e., for the subspaces that contain the ground state and the
lowest energy levels, $R=4$ for $M=1$, $R=2$ for $M=2, \dots, 8$
and 
then again $R=10$ for $8< M < 16$. The number of Lanczos iterations
for each random vector was determined by reaching convergence
for the two lowest energy levels. In subspaces with $M\ge 16$
the Hamiltonian was diagonalized completely.
The total computation time for
the \kagome\ system of 42 sites was $\sim 5\cdot 10^{6}$ core
hours at the Leibniz Supercomputing Center's \verb§supermuc§.

\section{Kagom\'e lattice antiferromagnet $N=42$}
\label{sec-3}

In what follows we focus on the specific heat, the density of
states, the uniform susceptibility, 
the entropy and the magnetization process.

\subsection{Zero-field properties}

\begin{figure}[ht!]
\centering
\includegraphics*[clip,width=0.8\columnwidth]{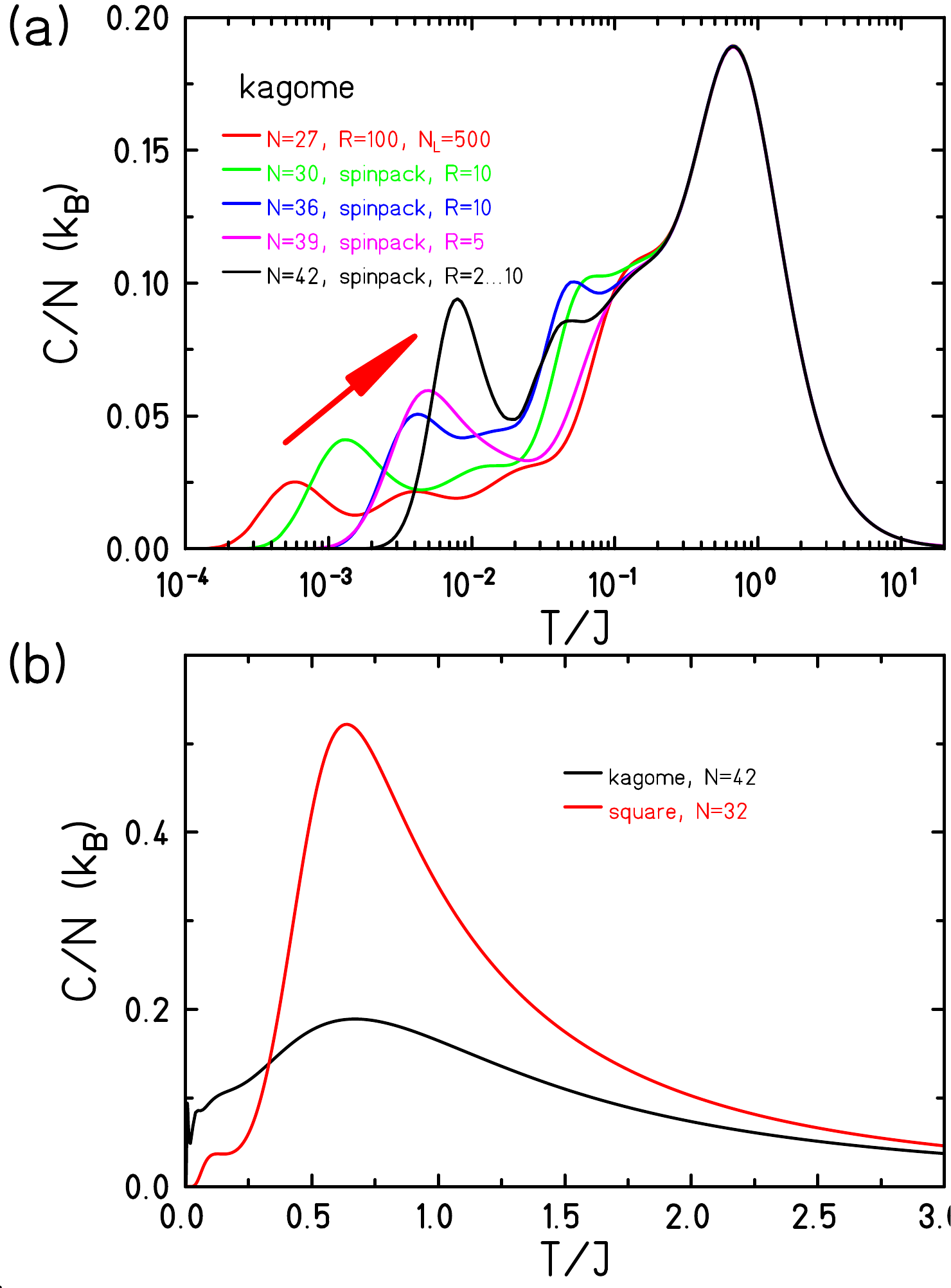}
\caption{(Color online) (a) Specific heat of the KHAF as function of temperature
  at $B=0$ for various systems sizes (logarithmic temperature
scale). (b) Specific heat of the KHAF and the SHAF as function of temperature
  at $B=0$ (linear temperature scale).}
\label{kago42-f-A}
\end{figure}

We start with the discussion of the specific heat $C(T)$, the
entropy $S(T)$ and the uniform 
susceptibility $\chi_0(T)$ using a logarithmic scale for $T$ in
order to make the 
low-temperature features transparent, see
Figs.~\ref{kago42-f-A}(a), \ref{kago42-f-C}(a), and \ref{kago42-f-D}(a).
The main maximum in the specific heat, set by the exchange
$J$, is at $T=0.67$, its height is
$C(T=0.67)/N=0.189$.\footnote{As is common praxis, temperatures
  and energies are given as multiples of $J$, thereby omitting
  $k_B$. $T=0.67$ thus means $k_B T=0.67 J$.}
Below $T=0.25$ the curvature of $C(T)$ changes and a shoulder-like profile
is present for $0.1 <T<0.25$.
This feature seems to be size-independent, i.e., finite-size effects 
appear likely only at $T<0.1$.
In this low-temperature region also a difference between odd and even
lattice sizes $N$ occurs, that is related to the GS value of
the total spin (doublet vs singlet), where even $N$ with a
singlet GS seem to better fit to the  
spin-liquid GS present for thermodynamically large systems.    

\begin{figure}[ht!]
\centering
\includegraphics*[clip,width=0.8\columnwidth]{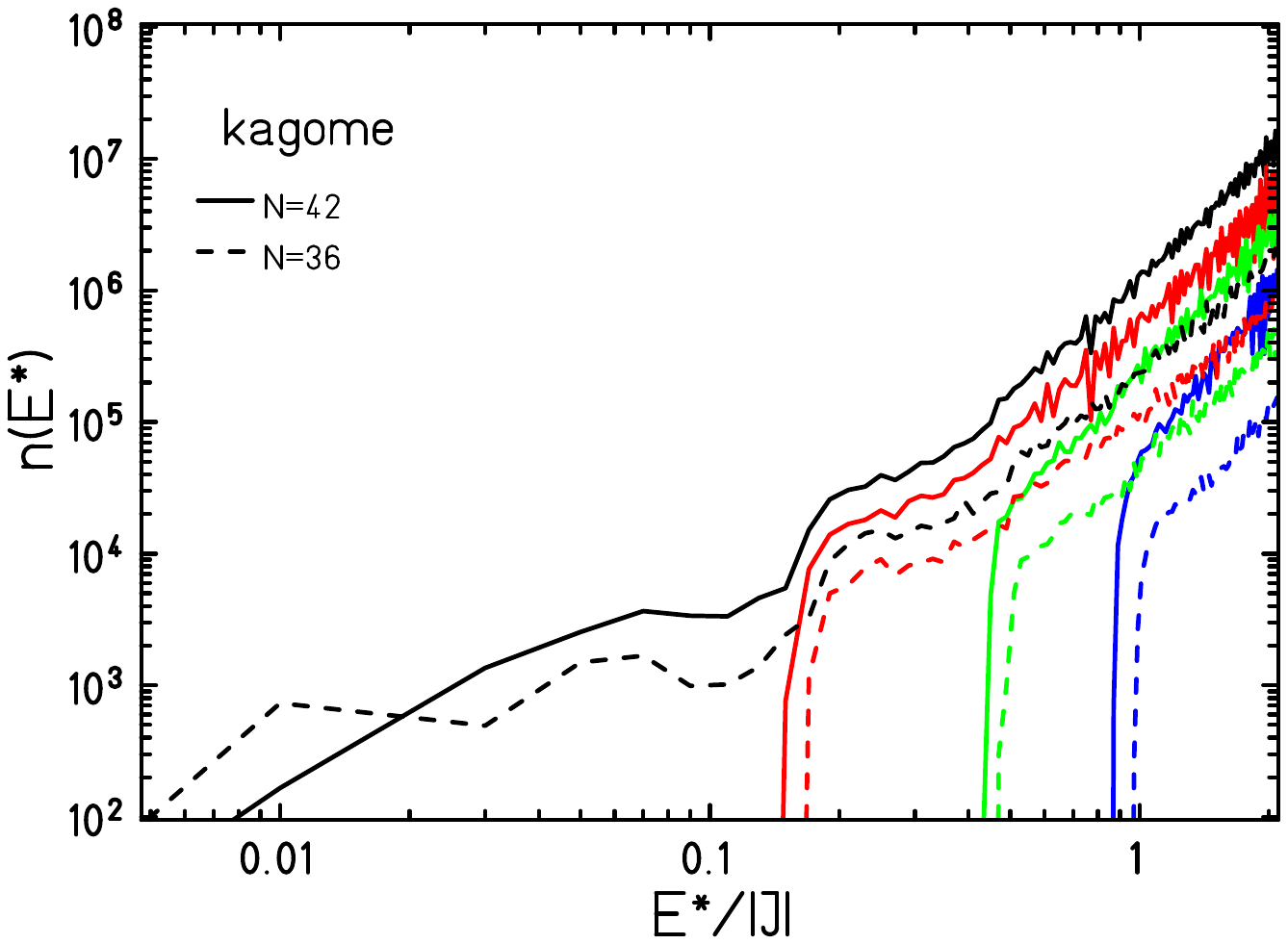}
\caption{(Color online) Binned density of states for $N=36$ (dashed
  curves) and $N=42$ (solid curves) as a function of the
  respective excitation energy $E^*$: total density of states --
  black, total density of states for $|M|=1$ -- red, for  $|M|=2$
  -- green, and for $|M|=3$ -- blue.}    
\label{kago42-f-B}
\end{figure}

\begin{figure}[ht!]
\centering
\includegraphics*[clip,width=0.8\columnwidth]{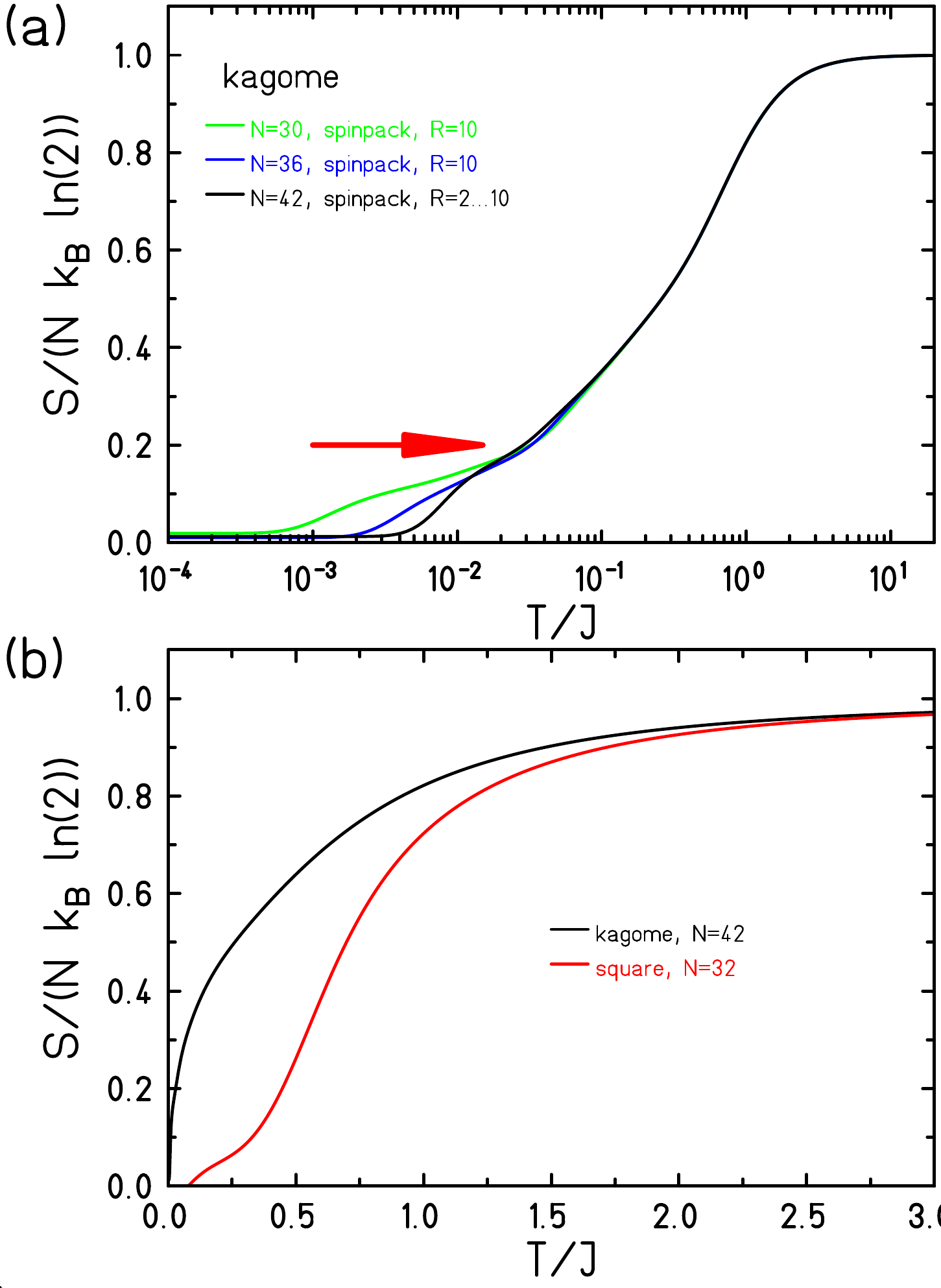}
\caption{(Color online) (a) Entropy of the KHAF as function of
temperature
  at $B=0$ for various systems sizes (logarithmic temperature
scale); the arrow marks the movement of the low-lying
density of states.
(b) Entropy  of the KHAF and the SHAF as function of temperature
  at $B=0$ (linear temperature  scale).}
\label{kago42-f-C}
\end{figure}

\begin{figure}[ht!]
\centering
\includegraphics*[clip,width=0.8\columnwidth]{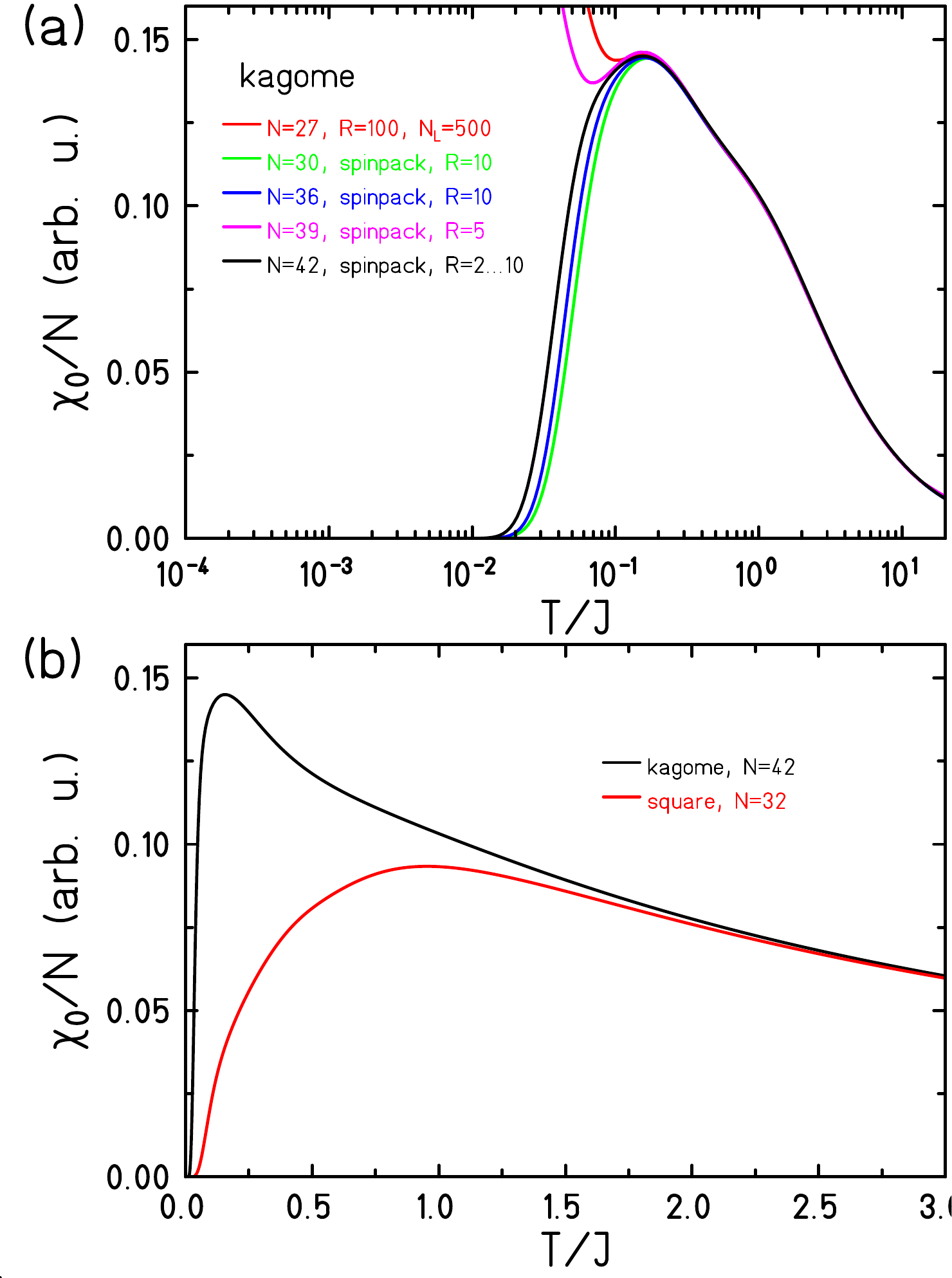}
\caption{(Color online) (a) Susceptibility of the KHAF as function of        
temperature at $B=0$ for various systems sizes (logarithmic
temperature scale).
(b) Susceptibility  of the KHAF and the SHAF as function of temperature       
  at $B=0$ (linear temperature scale).}
\label{kago42-f-D}
\end{figure}

The behavior at very low temperatures $T<0.1$ deserves a specific
discussion, where we focus on  even $N=30$, $36$ and $42$.
First we notice that for $N=36$ and $42$ just below the shoulder 
there is a rather flat maximum at about $T=0.05$. At very low
temperatures we observe a well pronounced extra peak in the
specific heat. This peak marks the appearance of low-lying
singlets above the ground state and is thus related to the
singlet-singlet gap. Common expectations are that such gaps
shrink with increasing size $N$. But in accordance with recent
exact diagonalization studies,\cite{Laeuchli2016} this peak moves
towards higher temperatures with increasing $N$, as highlighted
by the arrow in \figref{kago42-f-A}(a). One reason is that
the singlet-singlet gap as well as the singlet-triplet gap
do not shrink (considerably) when going from $N=36$ to $N=42$
and even $N=48$,
instead the singlet-singlet gap grows and the singlet-triplet
gap shrinks only slightly.\cite{Laeuchli2016}
This behavior can be further rationalized by looking at the
total density of states $n(E^*)$ as a function of the
respective excitation energy $E^*$
and the contributions of the different  
sectors of total magnetization $M$  to $n(E^*)$ as displayed in
\figref{kago42-f-B}. The density of states 
is evaluated by histograming the Krylov space energy eigenvalues
together with their respective weights. The bin  size is chosen as
$J/100$. From \figref{kago42-f-B} it becomes obvious which sector of
$M$ contributes to $C(T)/N$ at various low-temperature regimes.

Having in mind the sum rule 
\begin{equation} \label{sum-rule}
\int_0^\infty \frac{C(T)}{N T} dT =\int_{T=0}^{T=\infty} ds = s_\infty-s_0 =
k_B \log(2)
\ ,
\end{equation}
we may speculate that the weight of the extra peak at very low
temperatures moves towards the shoulder with increasing $N \to
\infty$, thus making the shoulder more pronounced. To conclude,
we argue that our results are in favor of a low-temperature
shoulder rather than an additional low-temperature maximum,
compare \figref{kago42-f-A}(b). This is in accordance with
recent tensor network calculations.\cite{Xi-Chen2017}

The behavior of the low-temperature peak also means that
concerning the density of singlet states, weight is not simply
shifted towards lower and lower energies with increasing $N$. It
may be that the singlet-singlet gap closes with increasing $N$,
but the density profile seems to behave differently, as can be
noted by comparing the dashed ($N=36$) and solid black ($N=42$)
curves in \figref{kago42-f-B}. This observation is further
supported by the behavior of the entropy $S(T)$ at $B=0$, which
is shown in \figref{kago42-f-C}. As  highlighted by the arrow,
the temperature above which the entropy rises moves towards
higher temperatures with increasing $N$ in accordance with the
motion of the low-temperature maximum of $C(T)$.

The singlet-triplet gap is even larger than the singlet-singlet
gap, therefore the zero-field susceptibility exhibits a gapped behavior,
as displayed in \figref{kago42-f-D}. For odd $N$ the ground
state possesses non-zero spin, therefore the susceptibility
diverges Curie-like in these cases. Since the singlet-triplet
gap does not move much with increasing size $N$, it is not
possible to draw definite conclusions about the functional form
of $\chi$ for $T\rightarrow 0$. Nevertheless, DMRG calculations
suggest that the singlet-triplet gap does not vanish in the
thermodynamic limit, but approaches $0.13(1)$.\cite{DMS:PRL12}

Finally, we compare
$C(T)$, $S(T)$ and $\chi_0(T)$ for the (highly frustrated) KHAF with the
corresponding FTLM data for the (unfrustrated) spin-$1/2$
square-lattice Heisenberg antiferromagnet 
(SHAF) of $N=32$ sites, see
Figs.~\ref{kago42-f-A}~(b), \ref{kago42-f-C}~(b), and \ref{kago42-f-D}~(b),
where we   use a linear temperature scale.
The temperature profile of all three quantities exhibits significant 
differences between the KHAF and the SHAF illustrating  the tremendous role of
frustration in a wide temperature range and, in particular, at low temperatures.\cite{MoR:PT06}
Note that at high temperatures
the quantities $C(T)$, $S(T)$ and $\chi_0(T)$ for both models approach each
other, since square and \kagome\ lattices 
have identical coordination number $z=4$.   
Thus the high-temperature series for 
$C$ and $\chi_0$ are identical up to order
$1/T^2$  and $1/T^3$, respectively, see,
e.g., Refs.~\onlinecite{Lohmann2014,Lohmann2011}.

\subsection{Field-dependent properties}

\begin{figure}[ht!]
\centering
\includegraphics*[clip,width=0.8\columnwidth]{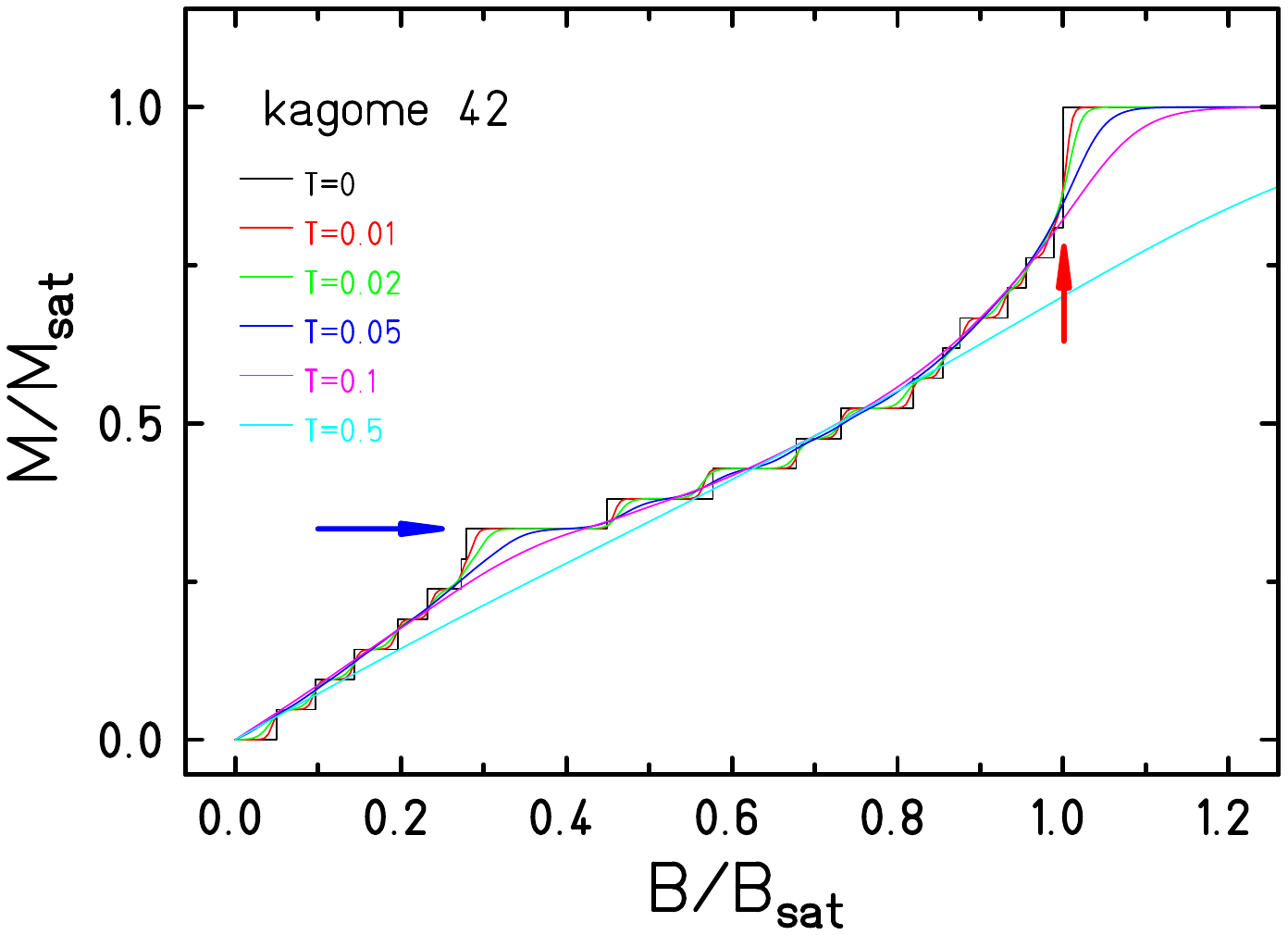}
\caption{(Color online) Magnetization vs applied magnetic field
  for various temperatures: both magnetization and field are
  normalized by their saturation values.}
\label{kago42-f-E}
\end{figure}
\begin{figure}[ht!]
\centering
\includegraphics*[clip,width=0.8\columnwidth]{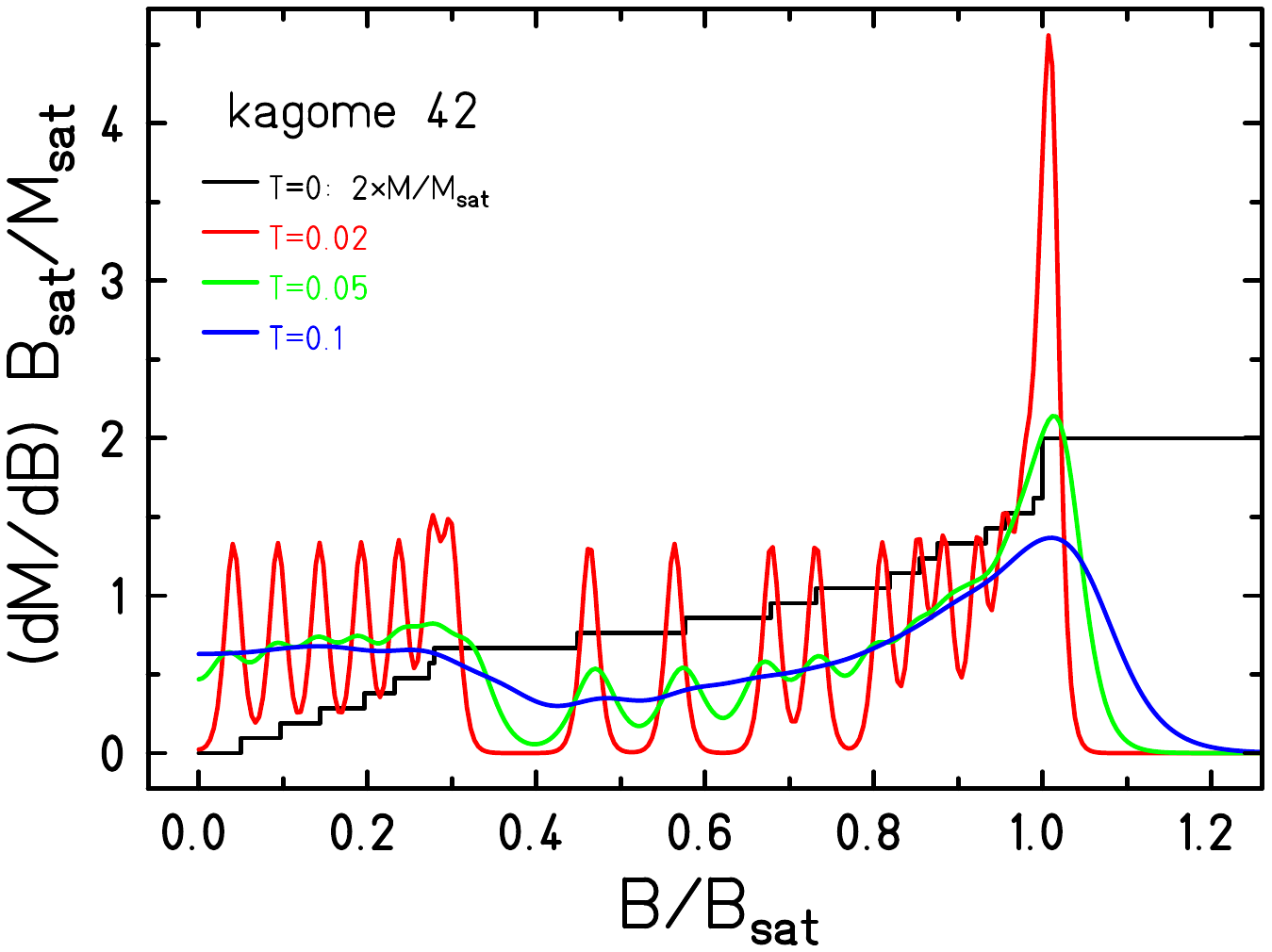}
\caption{(Color online) Derivative $dM/dB$ vs applied magnetic field $B$
  for various temperatures for $N=42$. Both, magnetization and field are
  normalized by their saturation values.}
\label{kago42-f-dMdB}
\end{figure}

The KHAF exhibits a number of
interesting properties in an applied magnetic
field.\cite{Hida2001,SHS:PRL02,HSR:JP04,HCG:PB05,SaN:PRB11, 
Nishimoto2013,CDH:PRB13,NaS:JPSJ14,SHS:PRL02,derzhko2004finite,zhitomirsky2004exact,NaS:JPSJ18}
Magnetization plateaus exist at $3/9(=1/3)$, $5/9$ and $7/9$ of the
saturation magnetization $M_{\rm sat}$ for the infinite system at $T=0$,
where the $1/3$ plateau is the widest. An additional
tiny plateau at
$1/9$ appears possible.\cite{Nishimoto2013}
Moreover, the magnetization curve at $T=0$ shows a macroscopic
jump to saturation due to the existence of independent localized
magnons.\cite{SSR:EPJB01,SHS:PRL02,derzhko2004finite,zhitomirsky2004exact,DRH:LTP07}  

In a calculation of a small lattice the magnetization curve is 
unavoidably a sequence of steps, that happen at ground state
level crossings at certain field values, compare
\figref{kago42-f-E}. 
Thus, due to this discretization the existence of 
smaller plateaus cannot be unambiguously deduced from such a
single magnetization curve.
Moreover, a specific plateau value $M_{\rm plateau}/M_{\rm sat}$
can be missed in the $M(B)$ curve, if it does not fit to the lattice size $N$.
For example, for our largest system of $N=42$ the values  at 
$M_{\rm plateau}/M_{\rm sat}= 5/9$ and $7/9$  are not present in
\figref{kago42-f-E} -- for $M(B)$ curves of other finite \kagome\ lattices, see,
e.g., Refs.~\onlinecite{HSR:JP04,CDH:PRB13,NaS:JPSJ14,NaS:JPSJ18}. 
Nevertheless, the major
plateau at $1/3$ (marked by the blue horizontal arrow) is
clearly visible in \figref{kago42-f-E}, since it is the widest
of all plateaus and it is very robust as a function of $N$.
\cite{SHS:PRL02,HSR:JP04,CDH:PRB13}
The jump to saturation (marked by the red vertical arrow) does not suffer from finite-size
effects. Its existence is analytically proven and persists for
all sizes.\cite{SSR:EPJB01,SHS:PRL02} 
We also mention that the pretty wide plateaus just above  
 the $1/3$-plateau most likely disappear for $N\to \infty$.
 \cite{Nishimoto2013}
The influence of the temperature on the $M(B)$ curve is relevant for 
experimental studies.
From \figref{kago42-f-E} it is obvious that for slightly
elevated temperatures the detection of plateaus by measuring $M(B)$ is difficult.
Therefore, the first
derivative  $dM/dB$ as a function of $T$ as presented in \figref{kago42-f-dMdB}
is often used in experiments to
find plateaus, cf., e.g., Ref.~\onlinecite{Tanaka_trian_2012}.
The $1/3$-plateau can be detected by the pronounced minimum  in
$dM/dB$.
Note that the oscillations of the red $dM/dB$ curve are also
due to finite-size effects.
It is worth mentioning that the position of the minimum  in
$dM/dB$ stemming from the $1/3$-plateau is shifted to higher
values of $B$ with increasing temperature.
Thus, for $T=0.05$ ($T=0.02$) it is at $B/B_{\rm sat}=0.398$ ($B/B_{\rm sat}=0.381$) whereas the midpoint of the
plateau is at $B/B_{\rm sat}=0.364$. For $T=0.1$ the minimum  in $dM/dB$ is
hardly detectable, see \figref{kago42-f-dMdB}.
This shift is related to the `asymmetric melting' of the plateau
due to
the larger density of low-lying excited
states below the plateau than that of low-lying excitations above the
plateau.\cite{Misawa2018}
Thus, for the KHAF the very existence of the plateau can be found 
by measuring $dM/dB$ at $T\lesssim 0.1$, but to determine the precise position of
it requires very low temperatures. Last but not least, we notice
that the jump of the magnetization to saturation at $T=0$ is washed out at $T>0$,
but its existence leads to a high peak in $dM/dB$ at the saturation field.

\begin{figure}[ht!]
\centering
\includegraphics*[clip,width=0.8\columnwidth]{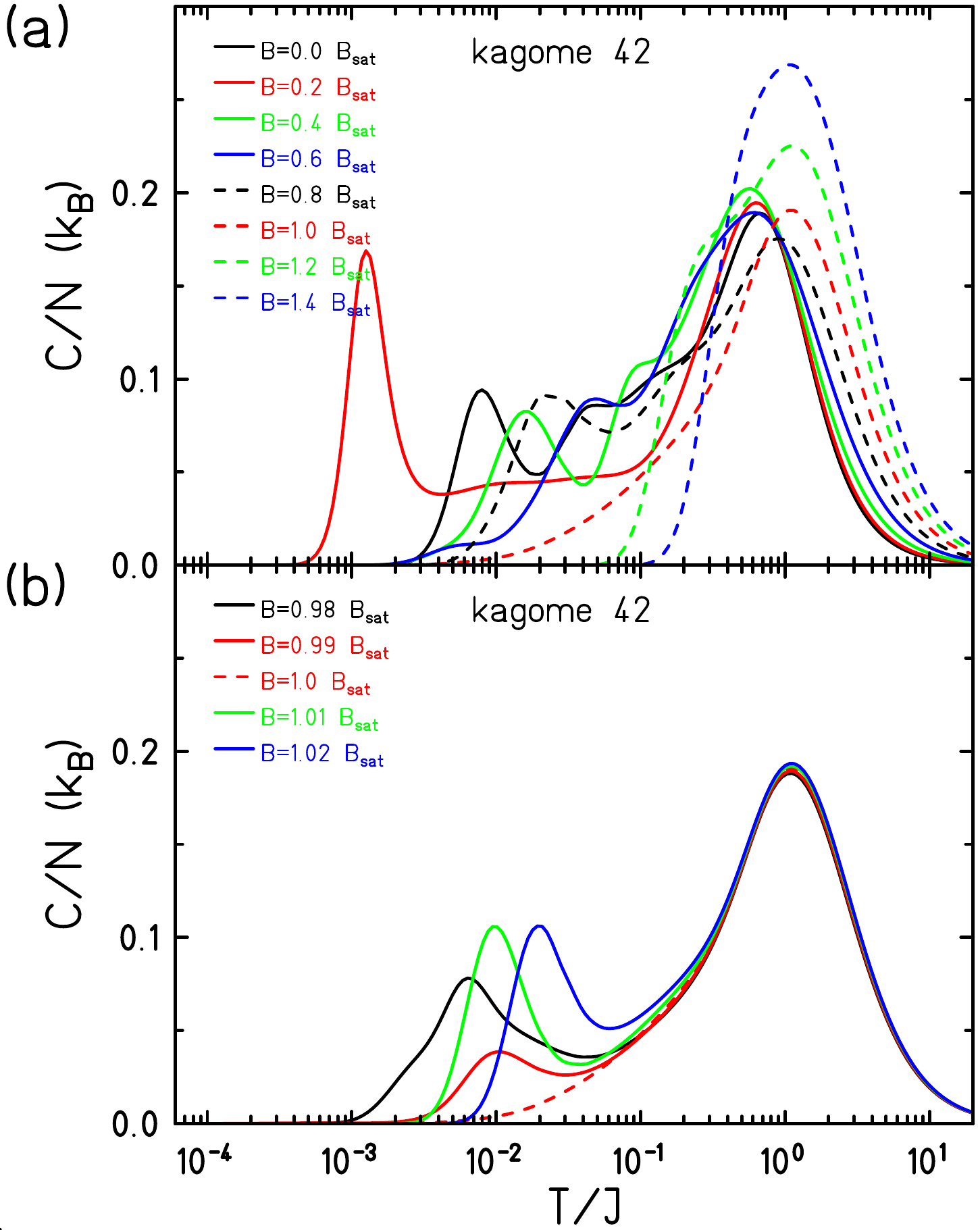}
\caption{(Color online) Specific heat of the KHAF as function of
  temperature for several values of the magnetic field varying from $B=0$ to $B=1.4B_{\rm sat}$
 (a) and for magnetic-field values at and close to the
  saturation field $B_{\rm sat}$ (b). 
}
\label{kago42-f-c}
\end{figure}

The influence of the magnetic field $B$ on the specific heat $C$ is shown in
\figref{kago42-f-c} for $N=42$. At very low temperatures and moderate fields the influence of
$B$ is determined by the shift of the low-lying magnetic excitations
with  $M=1$ and $M =2$ towards and even beyond the zero-field
singlet GS. As a result, the position and the height of 
the low-temperature (finite-size) peaks in $C(T)$ are substantially changed.
At temperatures below the main maximum there is no obvious systematic behavior
of $C(T)$ as a function of $B$, see \figref{kago42-f-c}(a). 
However, at magnetic fields slightly below and above the saturation field,  the
huge manifold of low-lying localized multimagnon
states (already mentioned in the introduction)
leads to an  
extra low-temperature maximum, see \figref{kago42-f-c}(b),
persisting in the thermodynamic
limit. \cite{SHS:PRL02,zhitomirsky2004exact,DRH:LTP07,DRM:IJMP15}   
It is worth mentioning, that for $B \lesssim B_{\rm sat}$  in the thermodynamic limit
this
extra-maximum likely becomes a true singularity indicating a
low-temperature order-disorder
transition into a magnon-crystal phase.\cite{zhitomirsky2004exact,RDK:PRB06}  

\begin{figure}[ht!]
\centering
\includegraphics*[clip,width=0.8\columnwidth]{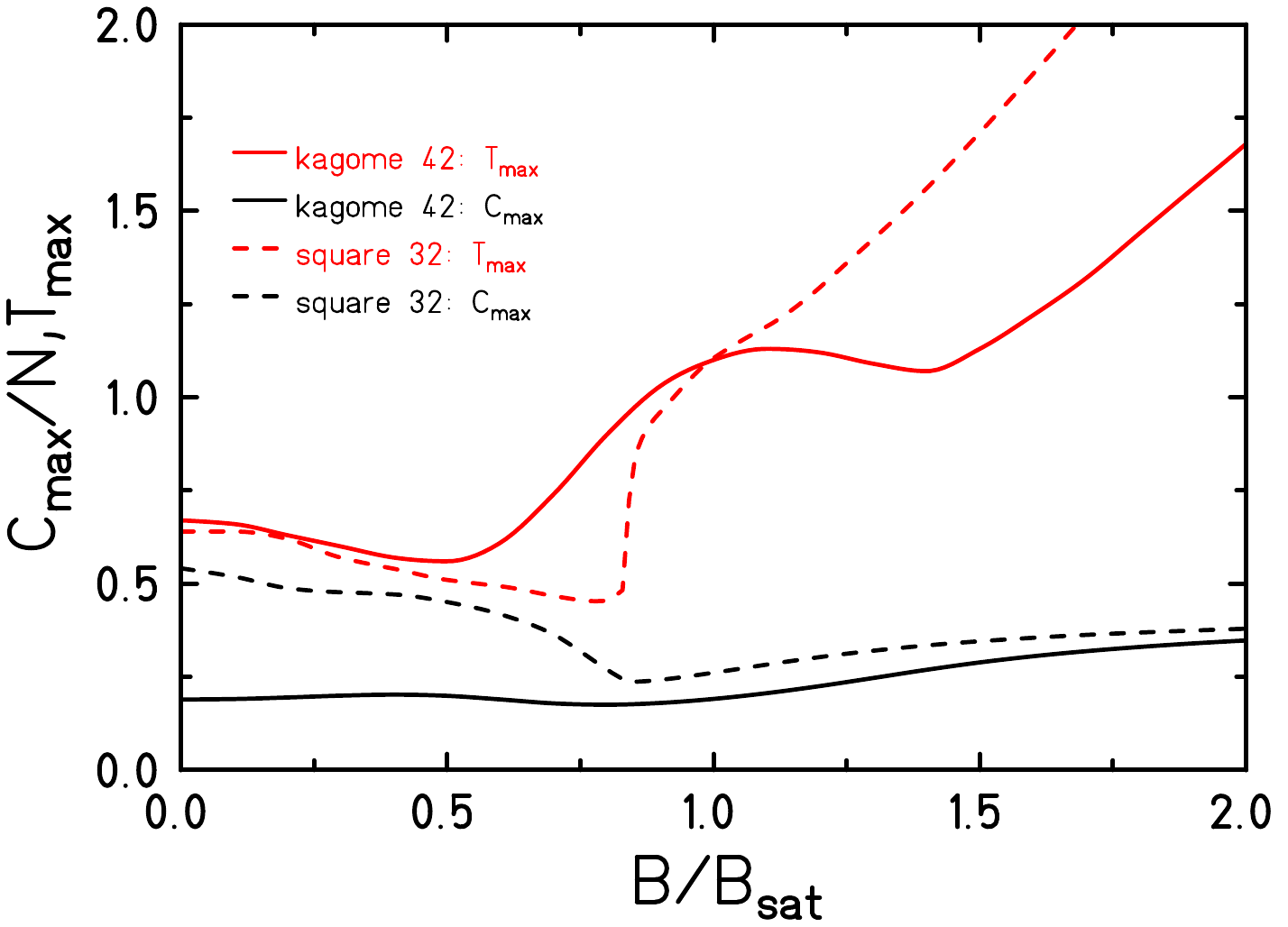}
\caption{(Color online) Position of the maximum in $C(T)$ in dependence of the
magnetic field $B$.}
\label{kago36-main-max}
\end{figure}

Interestingly, the influence of $B$ on the main maximum of $C$
depicted in \figref{kago36-main-max} shows some
systematics (see also Ref.~\onlinecite{Xi-Chen2017}, Fig.~3):
(i) The height of the
maximum $C_{\rm max}$ remains almost constant until  $B \sim 0.8
B_{\rm sat}$ and increases
smoothly for  $B >B_{\rm  sat}$. 
(ii) The position of the maximum $T_{\rm max}$ as a function of
$B$ exhibits two maxima at $B=0$ and $B \approx 1.1 B_{\rm
  sat}$ and two minima at   $B \approx 0.5 B_{\rm  sat}$ and $B
\approx 1.4 B_{\rm  sat}$ as well as two regions $0.65 B_{\rm
  sat} \lesssim B \lesssim 0.9 B_{\rm  sat}$ and $B \gtrsim 1.5
B_{\rm  sat}$ with an (almost) linear growth of $T_{\rm max}$.
To illuminate the role of frustration we contrast this behavior with that
of the  unfrustrated SHAF, also see \figref{kago36-main-max}.
At low magnetic fields the value of $T_{\rm max}$
is determined by $J$, and for both models $T_{\rm max}$ behaves very similar.
On the other hand, the difference in $C_{\rm max}$ is significant  and can be
related to the different low-energy physics which influences $C$ at higher
$T$ according to the sum rule (\ref{sum-rule}).
Beyond $B \sim 0.5 B_{\rm  sat}$ the different behavior is more evident.
The almost straight increase in $T_{\rm max}(B)$  
for $0.65 B_{\rm  sat} \lesssim B \lesssim 0.9 B_{\rm  sat}$ in
the case of the KHAF indicates a paramagnetic behavior.
The maximum around  $B = B_{\rm  sat}$ signals strong
frustration because it is related to the manifold 
${\cal W}$
of localized multi-magnon states setting an extra low-energy scale in the
vicinity of the saturation field in frustrated Heisenberg systems with
a flat  band.  For the specific flat-band model under consideration, i.e., the
KHAF, the number of localized multimagnon states grows
exponentially with system size as  ${\cal W} \sim
e^{0.111081N}$ and is thus relevant for the behavior of the
specific heat. \cite{zhitomirsky2004exact,derzhko2004finite}
The linear increase of $T_{\rm max}$ above $B_{\rm  sat}$
present in both models is then related to the paramagnetic
phase. Note, however, that this linear behavior starts at much
higher fields for the KHAF. We may therefore conclude,
that flat-band spin physics, that is typically relevant at
low-energy scales $T \ll J$, is
observable in the dependence of the specific heat on magnetic
field even at pretty high temperatures of $T \sim J$.

\section{Discussion}
\label{sec-4}

What is the gain of our new calculation for a KHAF with now
$N=42$ sites? 

First of all, it is by far the largest calculation of
thermodynamic properties such as magnetization and specific
heat. Other calculations such as Lanczos diagonalization for
$N=48$ as well as DMRG calculations aim at ground state
properties and at some low-lying states. Our results for $N=42$
and smaller sizes reveal that the specific heat very likely has
got a low-temperature shoulder instead of an additional
low-temperature maximum. 

As a second result we can show that `asymmetric melting' of the
1/3-plateau indeed occurs. Asymmetric melting influences our ability to
determine the plateau in measurements of $dM/dB$. We further
speculate that in addition to a non-balanced density of states
at the endpoints of the plateau the overall magnitude 
of the density of states grows with
the size $N$ of the lattice. This would mean that one would need
increasingly low temperatures in order to accurately measure
the 1/3-plateau.\cite{SaN:JKPS13,NaS:JPSJ14}

Further, we found that effects of strong frustration are not
only visible at low 
temperatures $T \ll J$, they are clearly visible at moderate (and high)
temperatures $ T \sim J$.
In particular, the very existence of a flat one-magnon band yielding a
huge manifold of low-lying localized multi-magnon states
leads to pronounced effects in the magnetization curve and
the temperature profile of the specific heat  at magnetic fields
near saturation.          

In accordance with Ref.~\onlinecite{Laeuchli2016}
we also found that in the low-field regime the convergence at
$T\lesssim 0.1 J$ to the thermodynamic limit is 
slow. Both the singlet-singlet gap as well as the
singlet-triplet gap change very little with increasing system
size. Therefore, at $B=0$ the behavior below
$T\lesssim 0.1 J$ seen in our calculations is still dominated by
finite-size effects.

Finally, we mention that the relation of our data to the experimental
data of the spin-liquid candidate
Herbertsmthite\cite{herbertsmithite2007,herbertsmithite2007a, 
Hiroi2009,herbertsmithite2009,herbertsmithite2010,herbertsmithite2012,CHF:PRL15}
is limited for several reasons.
First, the exchange interaction of this compound is estimated as
$J=190$~K. Low-temperature measurements of the magnetization as
well as the specific heat, in particular for $2 \text{K} \leq T \leq 10
\text{K}$, are well below $0.1 J$ and can thus, unfortunately, not
be compared with our simulations due the the finite-size effects below
$=0.1J$. Second, it is up to now
not settled how the interacting impurities as well as the
lattice vacancies in Herbertsmithite can be dealt with in
thermodynamic
calculations.\cite{HNW:PRB16,OBM:PRL17}
Moreover, there might be a noticeable 
spin anisotropy present in Herbertsmithite. Despite all the
uncertainties, it is believed that the spin liquid ground state
is a rather stable phenomenon.\cite{HNW:PRB16,SHH:PRB17,RHP:PRB17,OBM:PRL17,IPT:PRB18}

In view of the numerical effort of our investigations we
conjecture that exact diagonalization studies of the
thermodynamic behavior of the KHAF might be feasible for $N=45$
and $N=48$,\cite{Laeuchli2016} but larger systems must be dealt
with by, e.g., DMRG and tensor network
methods.\cite{Pollmann2017,Xi-Chen2017}

\section*{Acknowledgment}

This work was supported by the Deutsche
Forschungsgemeinschaft (DFG SCHN 615/23-1). Computing time at
the Leibniz Center in Garching is greatfully acknowledged.
JR is indebted to O.Derzhko for valuable discussions.


%

\end{document}